\documentclass[sigconf, authorversion, screen]{acmart}

\AtBeginDocument{%
  }

\copyrightyear{2025}
\acmYear{2025}
\setcopyright{rightsretained}
\acmConference[SIGCSE TS 2025]{Proceedings of the 56th ACM Technical Symposium on Computer Science Education V. 1}{February 26-March 1, 2025}{Pittsburgh, PA, USA}
\acmBooktitle{Proceedings of the 56th ACM Technical Symposium on Computer Science Education V. 1 (SIGCSE TS 2025), February 26-March 1, 2025, Pittsburgh, PA, USA}
\acmDOI{10.1145/3641554.3701782}
\acmISBN{979-8-4007-0531-1/25/02}

\makeatletter
\gdef\@copyrightpermission{
  \begin{minipage}{0.3\columnwidth}
   \href{https://creativecommons.org/licenses/by/4.0/}{\includegraphics[width=0.90\textwidth]{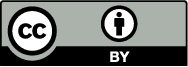}}
  \end{minipage}\hfill
  \begin{minipage}{0.7\columnwidth}
   \href{https://creativecommons.org/licenses/by/4.0/}{This work is licensed under a Creative Commons Attribution International 4.0 License.}
  \end{minipage}
  \vspace{5pt}
}
\makeatother

\usepackage[utf8]{inputenc} 
\usepackage{colortbl}
\usepackage{subcaption}
\usepackage{diagbox}
\usepackage{multirow}
\usepackage{xspace}
\usepackage{enumitem} 
\usepackage{mdframed}
\usepackage{booktabs} 
\usepackage{makecell} 
\usepackage[most]{tcolorbox} 
\tcbset{
  colframe=black, 
  colback=gray!10, 
  boxrule=0.2mm, 
  arc=2mm, 
  width=0.48\textwidth, 
  boxsep=0.75mm, 
  left=0.75mm, 
  right=0.75mm, 
  top=0.75mm, 
  bottom=0.75mm, 
  fontupper=\normalsize
}

\newcommand{\textcode}[1]{{\fontfamily{cmtt}\selectfont #1}\xspace}
\definecolor{darkgreen}{RGB}{0,100,0}
\definecolor{customblue}{HTML}{add8e6}
\definecolor{customdarkblue}{HTML}{4072a7}
\definecolor{customgreen}{HTML}{9aba56}
\definecolor{customorange}{HTML}{ed9943}
\definecolor{customred}{HTML}{ed4337}
\definecolor{customlightgreen}{HTML}{cdffcd}
\definecolor{customdarkgreen}{HTML}{00a64f}

\newcommand{\DSLRepeat}{\textcode{\textsc{Repeat}}}
\newcommand{\DSLRepeatUntil}{\textcode{\textsc{RepeatUntil}}}

\newcommand{\DSLIf}{\textcode{\textsc{If}}}
\newcommand{\DSLIfElse}{\textcode{\textsc{IfElse}}}

\newcommand{\currone}{\textsc{HoC-Fill}}
\newcommand{\currtwo}{\textsc{HoC-ACE}}

\newcommand{\currhoc}{\textsc{HoC}}
\newcommand{\curradv}{\textsc{PostHoC}}

\newcommand{\quizzes}{\text{quizzes}}

\definecolor{fontcellcolor}{RGB}{148,0,211}

\begin{document}

\title{Exploring the Impact of Quizzes Interleaved with Write-Code Tasks in Elementary-Level Visual Programming}

\author{Ahana Ghosh}
\affiliation{
  \institution{MPI-SWS}
  \city{Saarbr{\"u}cken}
  \country{Germany}  
}
\email{gahana@mpi-sws.org}

\author{Liina Malva}
\affiliation{
  \institution{MPI-SWS}
  \city{Saarbr{\"u}cken}
  \country{Germany}  
}
\affiliation{
  \institution{Tallinn University}
  \city{Tallinn}
  \country{Estonia}  
}
\email{liina.malva@tlu.ee}

\author{Alkis Gotovos}
\affiliation{
  \institution{MPI-SWS}
  \city{Saarbr{\"u}cken}
  \country{Germany}  
}
\email{agkotovo@mpi-sws.org}

\author{Danial Hooshyar}
\affiliation{%
  \institution{Tallinn University}
  \city{Tallinn}
  \country{Estonia}
}
\email{danial.hooshyar@tlu.ee}

\author{Adish Singla}
\affiliation{
  \institution{MPI-SWS}
  \city{Saarbr{\"u}cken}
  \country{Germany}  
}
\email{adishs@mpi-sws.org}

\renewcommand{\shortauthors}{Ahana Ghosh, Liina Malva, Alkis Gotovos, Danial Hooshyar, Adish Singla}
\renewcommand{\shorttitle}{Exploring the Impact of Quizzes within a Curriculum for Elementary-Level Visual Programming}

\begin{abstract}
We explore the role of quizzes in elementary visual programming domains popularly used for K-$8$ computing education. Prior work has studied various quiz types, such as fill-in-the-gap write-code questions. However, the overall impact of these quizzes is unclear: studies often show utility in the learning phase when enhanced with quizzes, though limited transfer of utility in the post-learning phase. In this paper, we aim to better understand the impact of different quiz types and whether quizzes focusing on diverse skills (e.g., code debugging and task design) would have higher utility. We design a study with \emph{Hour of Code: Maze Challenge} by code.org as the base curriculum, interleaved with different quiz types. Specifically, we examine two learning groups: (i) \currtwo{} with diverse quizzes including solution tracing, code debugging, code equivalence, and task design; (ii) \currone{} with simple quizzes on solution finding. We conducted a large-scale study with $405$ students in grades $6$--$7$. Our results highlight that the curriculum enhanced with richer quizzes led to higher utility during the post-learning phase.
\end{abstract}



\maketitle

\section{Introduction}\label{sec.intro}
\looseness-1Programming education at the K-$8$ levels has evolved to integrate inquiry-driven learning via quizzes alongside more traditional write-code tasks. Several popular platforms today such as the \emph{Express Curriculum}~\citep{codeorg_express} by code.org~\citep{codeorg}, \emph{ScratchJr}~\citep{velazquez2022designing}, and text-based programming (e)books~\citep{DBLP:conf/issep/RufBH15,DBLP:conf/iticse/Bower08a} incorporate microtasks in the form of quizzes to enhance student engagement, motivation, and provide formative feedback. Given the increasing adoption of quizzes in programming curricula, there is a need to assess their utility, in particular, how diverse quiz types impact learning and post-learning phases.

\looseness-1Prior work explored quizzes in several forms such as solution finding questions~\citep{DBLP:conf/aied/GhoshTDS22}, Parson's problems~\citep{DBLP:conf/icer/HouEW22}, and other microtasks such as code comprehension~\citep{DBLP:conf/sigcse/EricsonMD21,DBLP:journals/corr/abs-2401-01257}. Assessing the impact of these quizzes on learning is crucial to improve their design. However, studies have found the results to be mixed, often showing benefits during the learning phase but a limited transfer of learned skills during the post-learning (i.e., post-test) phase~\citep{DBLP:conf/sigcse/ZhiPMMBC19,DBLP:conf/icer/HassanZ21,DBLP:conf/icer/HouEW22,DBLP:conf/edm/PriceZB17}. 

\begin{figure*}
    \centering
    \begin{subfigure}{1\textwidth}
    \centering
    \includegraphics[width=1\textwidth]{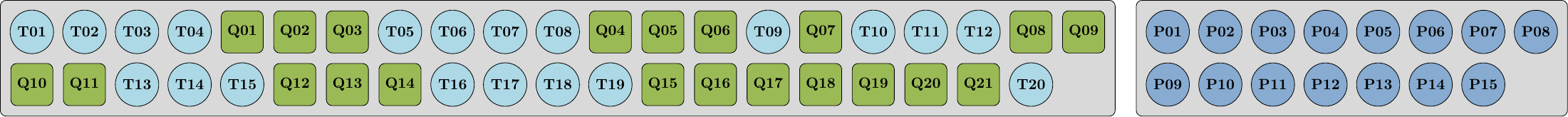}
    \caption{Items for \currtwo{} group. The quizzes, Q01--Q21, are based on solution-finding, code debugging, code equivalence, and task design.}
         \label{fig1:framework.hocace}
    \end{subfigure}
    \\
    \vspace{4mm}
    \begin{subfigure}{1\textwidth}
    \centering
     \includegraphics[width=1\textwidth]{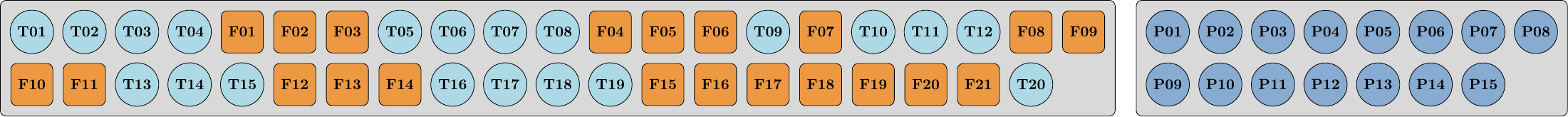}
    \caption{Items for \currone{} group. The quizzes, F01--F21, are fill-in-the-gap questions based on solution-finding.}
         \label{fig1:framework.hocfill}
    \end{subfigure}
    \caption{Sequence of items in the groups \currtwo{} and \currone{} during the learning and post-learning phases. The items in the left \textcolor{gray}{gray} box are used during the learning phase. In this set, the \textcolor{cyan!50}{light blue} circles containing ``T'' represent write-code tasks taken from \currhoc{} and the \textcolor{customgreen}{green} and \textcolor{customorange}{orange} squares represent quizzes. The items in the right \textcolor{gray}{gray} box are used during the post-learning phase. In this set, the \textcolor{customdarkblue}{dark blue} circles containing ``P'' represent write-code tasks taken from \curradv{}.
    }
\end{figure*}

\looseness-1In this work, we study the effect of different quiz types in introductory programming. We use a framework where write-code tasks taken from \emph{Hour of Code: Maze Challenge}~\citep{hourofcode_maze} by code.org~\citep{codeorg} (referred to as \currhoc{}) are interleaved with different quiz types. Specifically, we compare the utility of two sets of quizzes: one aligned with the three higher-order skills, \emph{analyzing, evaluating,} and \emph{creating} of Bloom's revised taxonomy of learning~\citep{anderson2001taxonomy}; the other set aligned with only the \emph{analyzing} level of the taxonomy and representing solution-finding questions found in literature~\citep{DBLP:conf/sigcse/EricsonMD21}. We evaluate their utility during the learning phase using \currhoc{} and during the post-learning phase on more advanced tasks (referred to as \curradv{}).  

\looseness-1To this end, we examine two learning groups: (i) the experimental group, \currtwo{}, where interleaved quizzes are based on solution tracing, code debugging/equivalence, and task design (Figure~\ref{fig1:framework.hocace}); (ii) the control group, \currone{}, where interleaved quizzes are based on fill-in-the-gap solution-finding questions (Figure~\ref{fig1:framework.hocfill}). We conduct a large-scale study with students in grades $6$--$7$ and center our analysis around the following research questions:  (1) \textbf{RQ1}: Does the experimental group outperform the control group during the learning phase on \currhoc{}? (2) \textbf{RQ2}: Does the experimental group outperform the control group during the post-learning phase on \curradv{}? (3) \textbf{RQ3}: How does the experimental group's performance compare to the control group considering students' prior experience?

\section{Related Work}\label{sec.relatedwork}
\looseness-1\paragraph{Types of quizzes.} Existing work has explored different types of inquiry-driven quizzes used in classrooms for teaching programming~\citep{DBLP:conf/ecai2/EneS19,DBLP:conf/iticse/SoltanpoorTD18,DBLP:journals/ijet/ZhangLZH20,DBLP:conf/issep/RufBH15,DBLP:conf/iticse/Bower08a,DBLP:conf/wipsce/RufBH13,maloney2010scratch,velazquez2022designing}. These include quizzes based on solution tracing posed as fill-in-gap questions~\cite {DBLP:conf/aied/GhoshTDS22} and Parson's problems to scaffold learners' write-code tasks~\citep{DBLP:conf/icer/HouEW22}. Recently, code.org~\citep{codeorg} introduced the \emph{Express Curriculum}~\citep{codeorg_express} containing microtasks based on code comprehension. In more complex text-based programming, the Runestone platform and Rust programming textbook used interactive quizzes alongside write-code tasks~\citep{DBLP:conf/sigcse/EricsonMD21,DBLP:journals/corr/abs-2401-01257}. Complementary to existing work, our work is focused on studying the impact of diverse quiz types in K-$8$ programming.

\looseness-1\paragraph{Quizzes in the context of learning taxonomies.} Various taxonomies have been proposed to categorize programming tasks based on problem-solving stages such as Loksa's six-stage problem-solving framework~\citep{DBLP:conf/chi/LoksaKJOMB16}, Hilton's ``Seven Steps''~\citep{DBLP:conf/comped/HiltonLR19}, and McCracken group's five-step approach~\citep{DBLP:journals/sigcse/McCrackenADGHKLTUW01}.  Programming quizzes can also be categorized and designed in the context of such stages~\citep{DBLP:conf/ace/PechorinaAD23,DBLP:conf/sigcse/LaymanWS07,DBLP:conf/iticse/Bower08a,DBLP:conf/issep/RufBH15}. A broader classification of programming quizzes can be done via Bloom's revised taxonomy of learning~\citep{anderson2001taxonomy}. It presents a five-level skill pyramid, emphasizing higher-order cognitive skills of \emph{analyzing}, \emph{evaluating}, and \emph{creating}. In this work, we adopt Bloom's revised taxonomy to classify programming quizzes, as it comprehensively covers these diverse skill levels as explored in~\citep{ghosh2024analyzing,padurean24benchmarking,DBLP:conf/ace/PorterC04}. 

\begin{figure*}[t!]
    \centering
    \scalebox{0.8}{
    \setlength\tabcolsep{10pt}
    \renewcommand{\arraystretch}{0.88}
        \begin{tabular}{ l  l  l  lll  l}
            \toprule
            \multicolumn{1}{l}{\textbf{Concept}} & 
            \multicolumn{1}{c}{\textbf{Tasks in \currhoc{}}} & 
            \multicolumn{1}{l}{\textbf{Tasks in \curradv{}}} & 
            \multicolumn{3}{c}{\textbf{Quizzes in \currtwo{}}} &  
            \multicolumn{1}{c}{\textbf{Quizzes in \currone{}}}\\ 
            \cmidrule(lr){4-6} \cmidrule(lr){7-7}
            {} & 
            \multicolumn{1}{c}{\textbf{}} & 
            \multicolumn{1}{c}{\textbf{}} & 
            \emph{\text{Analyzing}} & 
            \emph{\text{Evaluating}} & 
            \emph{\text{Creating}} & 
            \emph{\text{Analyzing}}\\
            \midrule
            {Basic moves and turns} & 
            {T01, T02, T03, T04, T05} & 
            {P01, P08}  & 
            Q01 & 
            Q02 & 
            Q03 & 
            F01, F02, F03 \\
            \cmidrule(lr){1-7}
            \DSLRepeat\textcode{\{\}} & 
            {T06, T07, T09} & 
            {P02} & 
            {Q04, Q05} & 
            Q07 & 
            Q06 & 
            F04, F05, F06, F07\\ 
            \DSLRepeat{}\textcode{\{\};}\DSLRepeat\textcode{\{\}} & 
            {T08} & 
            {}  & 
            {} & 
            {} & 
            {} & 
            {}\\
            \cmidrule(lr){1-7}
            \DSLRepeatUntil\textcode{\{\}} & 
            {T10, T11, T12, T13} & 
            {P03, P04, P09}  & 
            Q08 & 
            Q09 & 
            {Q10, Q11}  &  
            F08, F09, F10, F11\\
            \cmidrule(lr){1-7}
            \DSLRepeatUntil\textcode{\{}\DSLIf\textcode{\}} & 
            {T14, T15, T16, T17} & 
            {P05, P10}  & 
            Q12 &
            Q13 & 
            Q14  & 
            F12, F13, F14\\
            \cmidrule(lr){1-7}
            \DSLRepeatUntil\textcode{\{}\DSLIfElse{}\textcode{\}} & 
            {T18, T19} & 
            {P06} & 
             Q15 & 
            {Q16, Q19} & 
            {Q20, Q21}  & 
            F15, F16, F19, F20, F21\\
            \cmidrule(lr){1-7}
             \DSLRepeatUntil\textcode{\{}\DSLIfElse\textcode{\{\DSLIfElse{}\}\}} & 
             {T20} & 
             {P07, P15}  
             &  
             &  
             &   
             & \\
            \cmidrule(lr){1-7}
            \DSLRepeat{}\textcode{\{\};}\DSLRepeat{}\textcode{\{\};}\DSLRepeat{}\textcode{\{\}} & {} & 
            {P11}  
            &  
            &  
            &   
            & \\
            \DSLRepeat\textcode{\{\};}\DSLRepeatUntil\textcode{\{\}} & 
            {} & 
            {P12}  &  
            &  
            &   
            & \\
            \DSLRepeat\textcode{\{}\DSLRepeat\textcode{\}} & 
            {} & 
            {P13}  &  
            & 
            Q18 &   
            &  
            F18\\
            \DSLRepeat\textcode{\{}\DSLIf\textcode{\}} & 
            {} & 
            {P14}  & 
            Q17 & 
            &   
            & F17 \\
            \bottomrule
            \end{tabular}
    }
    \vspace{-2mm}
     \caption{
       \looseness-1 Distribution of items used in our study based on programming concepts of their solution codes. The quizzes are also classified based on higher-order skills of Bloom's revised taxonomy: \emph{analyzing, evaluating,  creating}~\citep{anderson2001taxonomy}. See Section~\ref{sec.ourcurr} for details.
    }
    \label{fig3:dist}
\end{figure*}

\begin{figure*}
    \centering
    \begin{subfigure}{0.48\textwidth}
    \centering
     \setlength{\fboxsep}{0.05pt}\fbox{\includegraphics[width=0.86\textwidth, trim=0 0.4cm 0 0, clip]{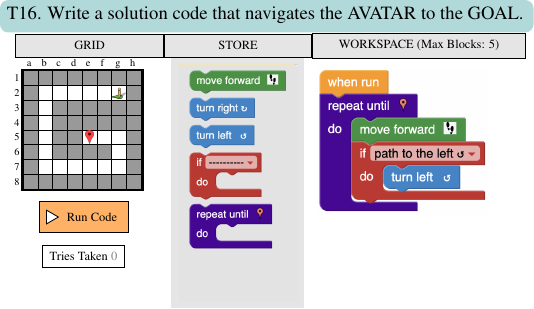}}
        \vspace{-2mm}
        \caption{Task T16 with its solution code}
        \label{fig1:example.hoc16}
    \end{subfigure}
    \begin{subfigure}{0.48\textwidth}
    \centering
     \setlength{\fboxsep}{0.05pt}\fbox{\includegraphics[width=0.86\textwidth, trim=0 0.4cm 0 0, clip]{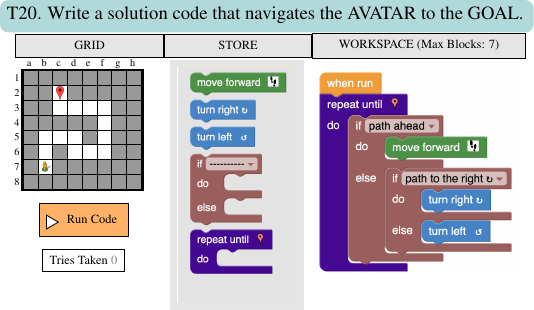}}
        \vspace{-2mm}
        \caption{Task T20 with its solution code}
        \label{fig1:example.hoc20}
    \end{subfigure}
    \\
     \vspace{2mm}
    \begin{subfigure}{0.48\textwidth}
    \centering
    \setlength{\fboxsep}{0.05pt}\fbox{\includegraphics[width=0.85\textwidth]{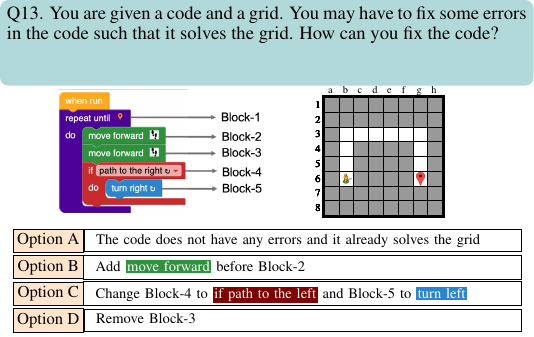}}
        \vspace{-2mm}
        \caption{Quiz Q13 from \currtwo{} (Solution: Option D)}
         \label{fig1:example.ace1}
    \end{subfigure}
    \begin{subfigure}{0.48\textwidth}
    \centering
    \setlength{\fboxsep}{0.05pt}\fbox{\includegraphics[width=0.85\textwidth]{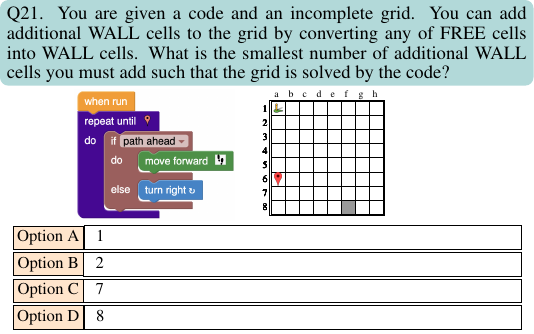}}
        \vspace{-2mm}
        \caption{Quiz Q21 from \currtwo{} (Solution: Option A)}
         \label{fig1:example.ace2}
    \end{subfigure}
    \\
    \vspace{2mm}
    \begin{subfigure}{0.48\textwidth}
    \centering
     \setlength{\fboxsep}{0.05pt}\fbox{\includegraphics[width=0.85\textwidth]{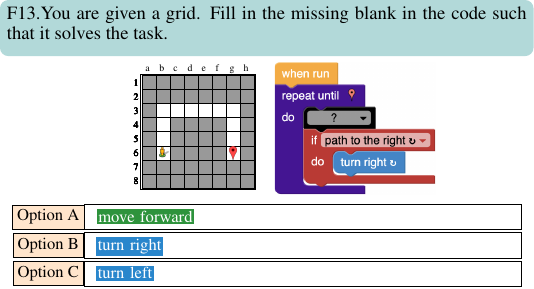}}
        \vspace{-2mm}
        \caption{Quiz F13 from \currone{} (Solution: Option A)}
         \label{fig1:example.hocfill1}
    \end{subfigure}
     \begin{subfigure}{0.48\textwidth}
    \centering
     \setlength{\fboxsep}{0.05pt}\fbox{\includegraphics[width=0.85\textwidth]{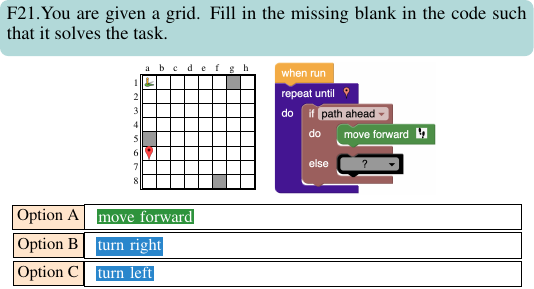}}
        \vspace{-2mm}
        \caption{Quiz F21 from \currone{} (Solution: Option B)}
         \label{fig1:example.hocfill2}
    \end{subfigure}
    \caption{
    \looseness-1(a) and (b) show examples of write-code tasks from \currhoc{}. (c) and (d) show quizzes based on code debugging and task design from \currtwo{}. (e) and (f) show corresponding quizzes based on solution-finding fill-in-the-gap questions from \currone{}.
    }
    \label{fig1:example}
    \vspace{-2mm}
\end{figure*}


\looseness-1\paragraph{Learning effects of quizzes} The impact of quiz types on learning has shown mixed results.  Generally, the effect of quizzes is evaluated in two stages: during the problem-solving process in the learning phase and through transfer effects in the post-learning phase. For instance, Parson's problems were found to be effective during the learning phase, but the effects did not translate in the post-learning phase~\citep{DBLP:conf/icer/HouEW22}. Fill-in-the-gap quizzes were also found to be effective during the learning phase but their effects in the post-learning phase were not measured~\citep{DBLP:conf/aied/GhoshTDS22}. In text-based programming, e.g., the Runestone platform and the Rust textbook~\citep{DBLP:conf/sigcse/EricsonMD21,DBLP:journals/corr/abs-2401-01257}, quizzes had a positive effect in the post-learning phase, with interactive quizzes promoting retention of concepts in (slightly) new contexts~\citep{perkins1992transfer}. In this work, we evaluate the utility of diverse quizzes both during the learning and post-learning phases in K-$8$ programming.

\section{Tasks and Quizzes in Our Study}\label{sec.ourcurr}
In this section, we describe the tasks and quizzes used in our study. First, we outline the popular programming curricula for K-$8$ learners, \emph{Hour of Code: Maze Challenge}~\citep{hourofcode_maze} by code.org~\citep{codeorg}, which forms the basis of our study design. Then, we describe the items for our learning groups, \currtwo{} and \currone{}. Finally, we present the \curradv{} tasks used in the post-learning phase.

\subsection{Items of \currhoc{}}\label{sec.ourcurr.hoc}
\looseness-1In this work, we use the popular block-based visual programming curriculum, \emph{Hour of Code: Maze Challenge}, as the basis for our study because of its simplicity in introducing programming concepts to novices. The tasks in this curriculum are write-code items where learners are provided with the following: a visual grid with an \textsc{Avatar}, a \textsc{Goal}, and \textsc{Wall}s; blocks of code; and a maximum limit on the number of blocks allowed. The objective is to write a solution code that when executed navigates the \textsc{Avatar} to the \textsc{Goal}, without crashing into any of the \textsc{Wall}s. We use all $20$ curriculum tasks in our study and refer to them as items T01--T20 (collectively called \currhoc{}). The distribution of these items by programming concepts is shown in Figure~\ref{fig3:dist}. Items T01--T08 are easier while T09--T20 are more difficult based on the complexity of their solution codes. Figures~\ref{fig1:example.hoc16} and \ref{fig1:example.hoc20} provide examples of items T16 and T20 with their solution codes.

\subsection{Items in the Two Learning Groups}\label{sec.ourcurr.groups}
\paragraph{\currtwo{}} Our experimental group, \currtwo{}, uses \currhoc{} curriculum interleaved with diverse quizzes based on \emph{analyzing, evaluating,} and \emph{creating} skills from Bloom's revised taxonomy~\citep{anderson2001taxonomy,bloom1956taxonomy} during the learning phase. For the quizzes, we selected multi-choice questions from the Computational Thinking assessment, \textsc{ACE}, specifically designed for K-$8$ learners~\citep{ghosh2024analyzing}. \textsc{ACE} includes a diverse set of $21$ quiz items based on solution tracing, code debugging, code equivalence, and task design. All $21$ items, referred to as Q01--Q21, are interleaved with the $20$ write-code tasks T01--T20 as shown in Figure~\ref{fig1:framework.hocace}.\footnote{Items T01--T20 are sequenced exactly as they occur in \currhoc{} which was in increasing complexity of their solution codes. The placement of interleaved quizzes was chosen to occur just after introducing programming concepts within the write-code tasks. The final sequence was decided based on feedback from K-$8$ teachers and a pilot study.} Their categorization based on the programming concepts and Bloom's revised taxonomy of skills is presented in Figure~\ref{fig3:dist}. Figures~\ref{fig1:example.ace1} and \ref{fig1:example.ace2} show two examples of quizzes in \currtwo{}.

\looseness-1\paragraph{\currone{}} Our control group, \currone{}, uses \currhoc{} curriculum interleaved with basic quizzes based on \emph{analyzing} skill of Bloom's revised taxonomy during the learning phase. Unlike \currtwo{}'s diverse quiz types, \currone{} only contains fill-in-the-gap solution-finding quizzes. These quizzes have $3$ answer options typically corresponding to the three basic actions. To maintain consistency with \currtwo{}, we designed $21$ quiz items for \currone{} (referred to as F01--F21) mirroring the task grids and solution codes of quizzes Q01--Q21. For example, the task grid and partial code in quiz F20 (Figure~\ref{fig1:example.hocfill2}), closely match quiz Q20 (Figure~\ref{fig1:example.ace2}). All $21$ quizzes are interleaved with the $20$ write-code tasks T01--T20 as shown in Figure~\ref{fig1:framework.hocfill}, occupying the same sequence positions as in \currtwo{}. Figures~\ref{fig1:example.hocfill1} and \ref{fig1:example.hocfill2} show two examples of quizzes in \currone{}.

\subsection{Items in \curradv{}}\label{sec.ourcurr.adv}
To evaluate the transfer effects for the two learning groups, we design a set of tasks in the post-learning phase called \curradv{}. These tasks are write-code tasks from the \emph{Hour of Code: Maze Challenge} domain, and include a combination of familiar write-code tasks from \currhoc{} as well as new write-code tasks that require applying a mix of programming concepts not encountered in \currhoc{}. Specifically, we design $15$ write-code items, referred to as P01--P15. Items P01--P07 are a subset of write-code tasks from \currhoc{} (specifically items T04, T07, T11, T13, T15, T18, T20), items P08--P15 are more challenging write-code tasks, and P10--P12 contain a combination of programming concepts unique to \curradv{}. The distribution of all $15$ items w.r.t. the programming concepts covered by their solution codes is shown in Figure~\ref{fig3:dist}.

\section{Study Setup and Demographics}\label{sec.study}
In this section, we outline our two-phase study setup, evaluation metrics, and demographics of the participating students.

\subsection{Phases of the Study}\label{sec.study.phases}
We design the study in two phases to assess the utility of quizzes in introductory programming during learning and post-learning. In the first phase, participants would engage with the learning material comprising the write-code tasks interleaved with quizzes specific to the learning group. In the second phase, participants from both groups would solve advanced write-code tasks to evaluate their post-learning performance.

We obtained an Institutional Review Board (IRB) approval from the Ethics Committee of Tallinn University before conducting the study. As part of the IRB approval, it was agreed that random allocation of participants for the two curriculum groups would be done at a class level and not at a student level to ensure consistency within a classroom. The study was conducted in Estonia where we pooled a total of $17$ different schools. Participation in the study was voluntary for teachers and students.

\begin{figure}[!t]
\centering
    \begin{minipage}{0.235\textwidth}
        \includegraphics[width=1\textwidth]{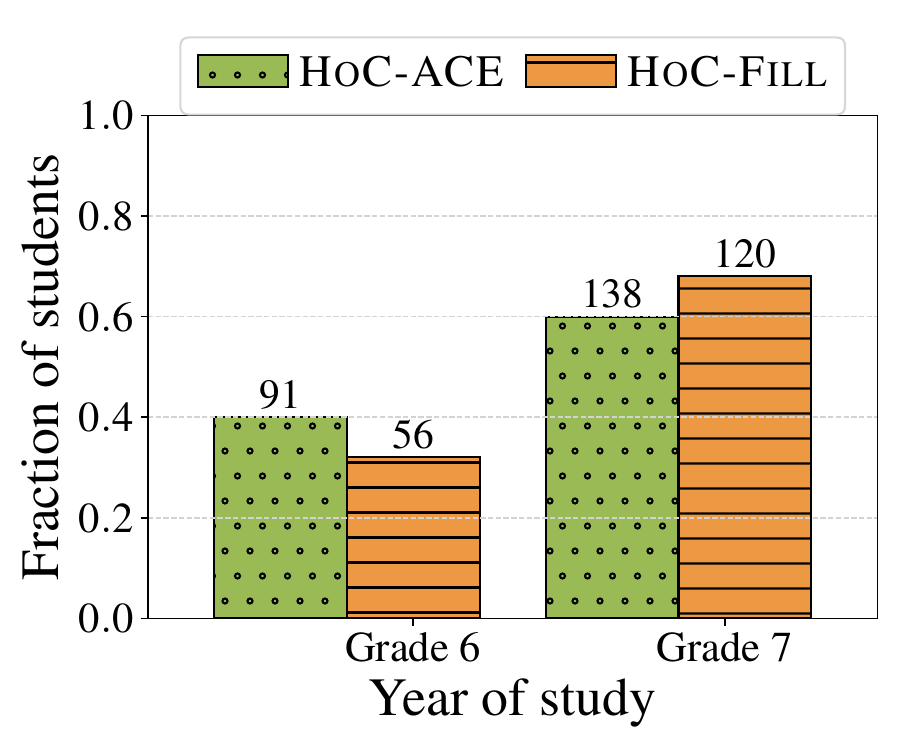}
    \end{minipage}
    \begin{minipage}{0.235\textwidth}
         \includegraphics[width=1\textwidth]{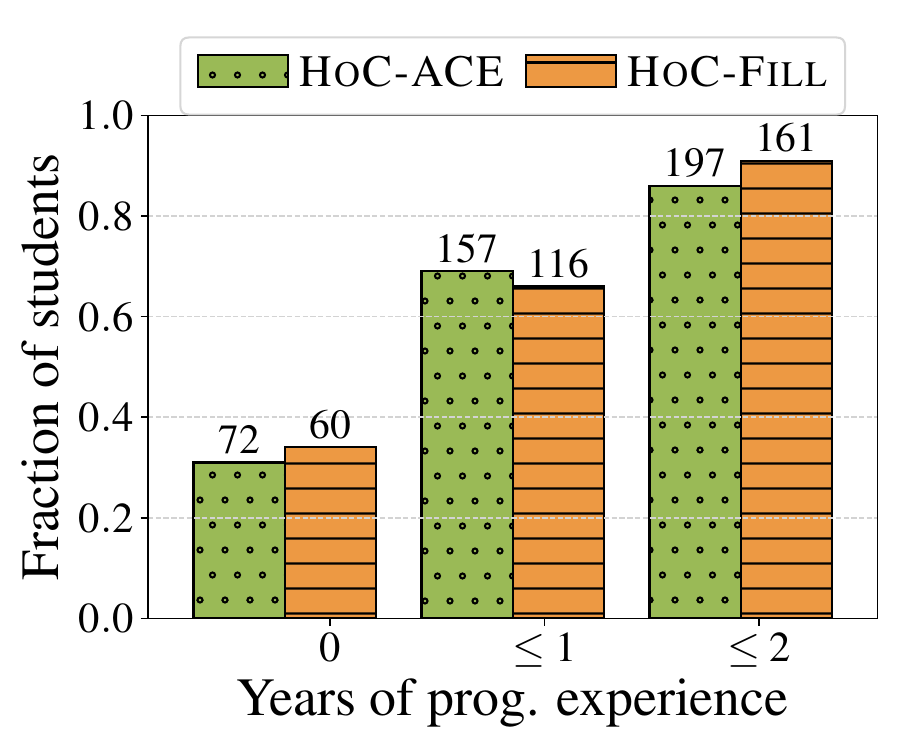}
    \end{minipage}
    \caption{\looseness-1Distribution of students' grade (year of study) on the left and their self-reported years of programming experience on the right based on the pre-survey data collected. The fraction of students is computed w.r.t. each learning group.}
    \label{fig:priorprogexp}
\end{figure}
\begin{figure*}[!t]
\centering
    \begin{subfigure}{0.23\textwidth}
    \centering
        \includegraphics[width=1\textwidth, trim=0 0.28cm 0 0, clip]{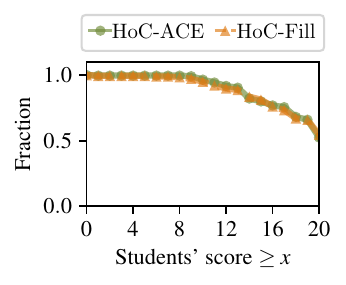}
        \caption{Score distribution}
        \label{fig4:overall.hocmaze}
    \end{subfigure}
    \begin{subfigure}{0.74\textwidth}
    \centering
        \scalebox{0.90}{
        \setlength\tabcolsep{4pt}
        \renewcommand{\arraystretch}{1.45} 
        \begin{tabular}{c cc cl c c c c}
                \toprule
                {} & 
                \multicolumn{2}{c}{\textbf{Performance}} & 
                \multicolumn{2}{c}{\textbf{Time (s)}} &
                {} &
                \multicolumn{3}{c}{\textbf{Performance per item category}}
                \\
                \cmidrule(lr){2-3} \cmidrule(lr){4-5} \cmidrule(lr){7-9}
                {} & 
                \small{{T01--T20}} & 
                \small{{\quizzes{}}} & 
                \small{{T01--T20}} & 
                \small{{\quizzes{}}} &
                {} &
                \small{{T01--T08}} &
                \small{{T09--T20}} &
                \small{{T04, T07, T11, T13,}}
                \\
                 {} & 
                 {} & 
                 {} & 
                 {} &
                 {} & 
                 {} &
                 {} &
                 {} &
                 \small{T15, T18, T20}
                 \\
                \midrule
                 \textbf{\small{{\currtwo{}}}}  &  
                     $0.885 \ \textcolor{gray}{(0.01)}$ & 
                     $0.931 \ \textcolor{gray}{(0.01)}^{\textcolor{blue}{\ast}}$ & 
                     $126\ \textcolor{gray}{(2.8)}$ & 
                     $32\ \textcolor{gray}{(1.0)}^{\textcolor{blue}{\ast\ast}}$ &
                     {} &
                     $0.993 \ \textcolor{gray}{(0.00)}$ & 
                     $0.813 \ \textcolor{gray}{(0.02)}$ & 
                     $0.842 \ \textcolor{gray}{(0.01)}$
                     \\
                    \textbf{\small{\currone{}}} & 
                    $0.879 \ \textcolor{gray}{(0.01)}$ & 
                    $0.917 \ \textcolor{gray}{(0.01)}$ & 
                    $124\ \textcolor{gray}{(3.5)}$ & 
                    $18 \ \textcolor{gray}{(0.4)}$ &
                    {} &
                    $0.987 \ \textcolor{gray}{(0.01)}$ & 
                    $0.806 \ \textcolor{gray}{(0.02)}$ & 
                    $0.832 \ \textcolor{gray}{(0.02)}$
                    \\
                     \bottomrule
        \end{tabular}
        }
        \caption{Aggregated and item-wise performance}
        \label{fig4:overall.aggr}
    \end{subfigure}
    \vspace{-1mm}    
    \caption{\looseness-1Performance in the learning phase. \textbf{(a)} shows the fraction of students who scored above a certain threshold on \currhoc{} (with maximum score $20$). \textbf{(b)} shows the  mean \textcolor{gray}{(std err)} performance in aggregate and per item category scaled between $0$ and $1$. The ``Time (s)'' column shows the average seconds spent per item. $^{\textcolor{blue}{\ast\ast}}$ indicates significance with $p < 0.01$ and $^{\textcolor{blue}{\ast}}$ with $p < 0.05$.
    }
    \label{fig4:overall}
    \vspace{-1mm}
\end{figure*}

Data was collected between December 2023 and March 2024. During both phases, a participant received the same unique username to ensure anonymity. The first phase involved two $45$-minute lessons over two weeks. Participants begin by completing a brief pre-survey on demographics and programming pre-knowledge; then they solved items in their assigned learning group on a dedicated web platform, with multiple attempts allowed per task and quiz. This phase had a maximum score of $20$ points, measured on the write-code tasks from \currhoc{} ($1$ point per task), and is referred to as their learning phase performance. The second phase occurred in the third week and included a $45$-minute session where participants first completed a brief post-survey assessing their learning experience; then they solved advanced write-code tasks of \curradv{}. This phase had a maximum of $15$ points, measured on \curradv{} ($1$ point per task), and is referred to as their post-learning performance.

\subsection{Participating Students and Experience}\label{sec.study.preknowledge}
\looseness-1A total of $405$ students (between the ages of $10$--$15$) from $37$ classes across $17$ schools participated in the study with the following distribution w.r.t. grade and gender: $147$ students from grade 6 and $258$ students from grade 7; $170$ female students and $235$ male students. The distribution of classes was random across both learning groups, and their counts are as follows: $18$ classes in \currone{} ($n = 176$ students), and $19$ classes in \currtwo{} ($n = 229$ students).

\looseness-1We used pre-survey data to gather students' programming experience and preknowledge for any discrepancies between the two learning groups. Only about $3\%$ of students in each group were familiar with \emph{Hour of Code: Maze Challenge}.

Figure~\ref{fig:priorprogexp} shows the distribution of students' grade (year of study) and their reported number of years of prior programming experience. A $\chi^2$ significance analysis~\citep{cochran1952chi2} revealed no significant differences between the groups w.r.t. grade ($p > 0.12$) and w.r.t. prior programming experience ($p > 0.78$), indicating that the two groups were similar in terms of academic and programming experience.

\subsection{Collected Data and Analysis Setup}\label{sec.study.analysis.setup}
After our data collection, we processed the following information for each phase of the study for analysis of our RQs:
\begin{itemize}[left=0pt]
    \item \emph{Learning phase}: We analyzed the success, attempts, and time taken on \currhoc{} tasks (T01--T20) for both groups as well as on quizzes Q01--Q21 for \currtwo{} and F01--F21 for \currone{}.
    \item\emph{Post-learning phase}: We analyzed the success, attempts, and time taken on \curradv{} tasks (P01--P15) for both groups.
\end{itemize}

\looseness-1Our analysis for RQs in Section~\ref{sec.results} is primarily focused on comparing the two learning groups in the post-learning phase on \curradv{}. To further evaluate the relative utility of these groups, we collected additional statistics on \curradv{} from students who did not engage with any quizzes when solving the standard \currhoc{} tasks. For this purpose, we requested teachers across $14$ classes (different from the $37$ classes used in the above data collection) to report on students' performance on \curradv{}, after they have completed the standard \currhoc{} curriculum in one $45$-minute lesson without quizzes. We present a relative comparison w.r.t. this baseline performance in Section~\ref{sec.results.rq2}.

\begin{figure*}[!t]
\centering
    \begin{subfigure}{0.23\textwidth}
    \centering
        \includegraphics[width=4.1cm, trim=0 0.28cm 0 0, clip]{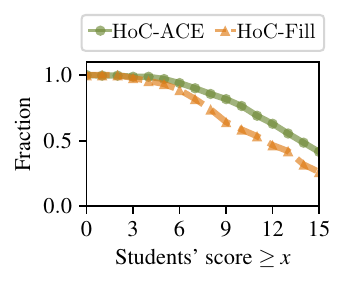}
        \caption{Score distribution}
        \label{fig5:overall.hocmaze}
    \end{subfigure}
    \begin{subfigure}{0.76\textwidth}
    \centering
        \scalebox{0.90}{
        \setlength\tabcolsep{2pt}
        \renewcommand{\arraystretch}{1.6} 
        \begin{tabular}{cc l c c c l c l c l}
                \toprule
                {} & 
                {} &
                \multicolumn{1}{c}{\textbf{Performance}} & 
                {} &
                \multicolumn{1}{c}{\textbf{Time(s)}} &
                {} &
                \multicolumn{5}{c}{\textbf{Performance per item category}}
                \\
                \cmidrule(lr){3-3} \cmidrule(lr){5-5} \cmidrule(lr){7-11}
                {} & 
                {\ \ \ \ \ } &
                \multicolumn{1}{c}{P01--P15} & 
                {\ \ \ \ \ } &
                \multicolumn{1}{c}{P01--P15} &
                {\ \ \ \ } &
                \multicolumn{1}{c}{P01--P07} &
                {\ \ \ \ } &
                \multicolumn{1}{c}{P08--P15} &
                {\ \ \ \ } &
                \multicolumn{1}{c}{P10--P12}
                \\
                \midrule
                \textbf{\small{{\currtwo{}}}}  &  
                 {\ \ \ \ \ \ \ \ \ \ } &
                 $0.799\, \textcolor{gray}{(0.02)}^{\textcolor{blue}{\ast\ast}}$ & 
                 {\ \ \ \ \ \ \ \ \ \ } &
                 $122 \, \textcolor{gray}{(1.9)}$ &
                 {\ \ \ \ \ \ \ \ \ \ } &
                 $0.890 \, \textcolor{gray}{(0.01)}^{\textcolor{blue}{\ast\ast}}$ & 
                 {} &
                 $0.719 \, \textcolor{gray}{(0.02)}^{\textcolor{blue}{\ast\ast}}$ & 
                 {} &
                 $0.742 \, \textcolor{gray}{(0.02)}^{\textcolor{blue}{\ast\ast}}$
                 \\
                \textbf{\small{\currone{}}} & 
                {\ \ \ \ \ \ \ \ \ \ } &
                $0.702 \, \textcolor{gray}{(0.02)}$ & 
                {\ \ \ \ \ \ \ \ \ \ } &
                $119 \, \textcolor{gray}{(2.8)}$ &
                {\ \ \ \ \ \ \ \ \ \ } &
                $0.843 \, \textcolor{gray}{(0.01)}$ & 
                {} &
                $0.579 \, \textcolor{gray}{(0.03)}$ & 
                {} &
                $0.593 \, \textcolor{gray}{(0.03)}$
                \\
                 \bottomrule
        \end{tabular}
        }
        \caption{Aggregated and item-wise performance}
        \label{fig5:overall.aggr}
    \end{subfigure}
    \caption{\looseness-1Performance in the post-learning phase. \textbf{(a)} shows the fraction of students who scored above a certain threshold on \curradv{} (with maximum score $15$). \textbf{(b)} shows the  mean \textcolor{gray}{(std err)} performance in aggregate and per item category scaled between $0$ and $1$ for each group. The ``Time (s)'' column shows the average seconds spent per item. $^{\textcolor{blue}{\ast\ast}}$ indicates significance with $p < 0.01$.
    }
    \label{fig5:overall}
\end{figure*}
\begin{figure*}[!h]
\centering
        \scalebox{0.90}{
        \setlength\tabcolsep{6.5pt}
        \renewcommand{\arraystretch}{1.3}         
        \begin{tabular}{c ll lll rr rrr }
                \toprule
                & \multicolumn{5}{c}{\textbf{Performance on \curradv{}}} & \multicolumn{5}{c}{\textbf{Participant Count}}\\
                \cmidrule(lr){2-6} \cmidrule(lr){7-11}
                & \multicolumn{2}{c}{{Grade}} & \multicolumn{3}{c}{{Years of prog. experience}} & \multicolumn{2}{c}{{Grade}} & \multicolumn{3}{c}{{Years of prog. experience}}\\
                \cmidrule(lr){2-3} \cmidrule(lr){4-6} \cmidrule(lr){7-8} \cmidrule(lr){9-11}
                & \multicolumn{1}{c}{Grade 6} &  \multicolumn{1}{c}{Grade 7} &  \multicolumn{1}{c}{$0$} &  \multicolumn{1}{c}{$\leq 1$} & \multicolumn{1}{c}{$\leq 2$} & \multicolumn{1}{c}{Grade 6} &  \multicolumn{1}{c}{Grade 7} &  \multicolumn{1}{c}{$0$} &  \multicolumn{1}{c}{$\leq 1$} & \multicolumn{1}{c}{$\leq 2$}\\
                 \hline
                 \textbf{\small{{\currtwo{}}}}  & $0.808 \, \textcolor{gray}{(0.02)} ^{\textcolor{blue}{\ast\ast}}$ 
                 & $0.793 \, \textcolor{gray}{(0.02)}^{\textcolor{blue}{\ast\ast}}$ 
                 & $0.810 \, \textcolor{gray}{(0.03)}^{\textcolor{blue}{\ast\ast}}$ 
                 & $0.808 \, \textcolor{gray}{(0.02)}^{\textcolor{blue}{\ast\ast}}$ 
                 & $0.805 \, \textcolor{gray}{(0.02)}^{\textcolor{blue}{\ast\ast}}$
                 & $91$ & $138$ & $72$ & $157$ & $197$
                 \\
                \textbf{\small{\currone{}}} & $0.650 \, \textcolor{gray}{(0.03)}$ 
                & $0.727 \, \textcolor{gray}{(0.02)}$ 
                & $0.716 \, \textcolor{gray}{(0.04)}$ 
                & $0.706 \, \textcolor{gray}{(0.02)}$ 
                & $0.692 \, \textcolor{gray}{(0.02)}$ 
                & $56$ & $120$ & $60$ & $116$ & $161$\\
                 \bottomrule
        \end{tabular}
        }
    \caption{\looseness-1 Performance on \curradv{} w.r.t experience. Aggregated mean \textcolor{gray}{(std err)} performance scaled between $0$ and $1$ on \curradv{} based on students' grade (year of study) and years of prior programming experience. $^{\textcolor{blue}{\ast\ast}}$ indicates significance with $p < 0.01$.
    }
    \label{fig6:overall}
\end{figure*}

\section{Results and Discussion}\label{sec.results}
In this section, we discuss the results of our study centered around the research questions (RQs) introduced in Section~\ref{sec.intro}.

\subsection{RQ1: Performance in Learning Phase}\label{sec.results.rq1}
\looseness-1To address RQ1, we assess the performance of the groups in the learning phase on \currhoc{} and the quizzes. The distribution of scores on \currhoc{} for each learning group is shown in Figure~\ref{fig4:overall.hocmaze}. In Figure~\ref{fig4:overall.aggr}, we present the normalized aggregated performance (scaled between $0$ and $1$) of both groups on the learning phase items, write-code tasks (T01--T20), and quizzes. A $\chi^2$ significance analysis~\citep{cochran1952chi2} showed no significant difference in aggregate performance ($\chi^2 = 0.64, p = 0.42$) between \currone{} and \currtwo{}. $\chi^2$ analysis w.r.t. each item category also showed no significant differences (all $p$-values $>0.14$). We also analyzed the time spent by each learning group on the write-code tasks and quizzes using Welch's $t$-test of significance~\cite{welch1947generalization}. While both groups spent similar time on write-code tasks, \currtwo{} group spent double the time on quizzes than the \currone{} group, and this difference was significant ($t = 11.9, p < 0.01$). 

\looseness-1Our analysis highlights that both groups performed similarly in the learning phase suggesting that the type of quizzes did not significantly impact overall performance during this phase. Interestingly, based on the time analysis we can conclude that students spent significantly longer time solving the quizzes in \currtwo{} compared to the quizzes in \currone{}. Investigating this further, we looked at the average attempts on the quizzes by both groups and they were as follows: $1.88$ for \currtwo{} and $1.32$ for \currone{}. Furthermore, the success rate on the quizzes within the first attempt for the groups was as follows: $0.849$ for \currtwo{} and $0.880$ for \currone{}, indicating the relative difficulty of the quizzes. However, in subsequent attempts, the success rate increased to $0.931$ for \currtwo{} and $0.917$ for \currone{}.

\begin{tcolorbox}
\textbf{Answer to RQ1}: The experimental group and the control group perform similarly in the learning phase on \currhoc{}.
\end{tcolorbox}

\subsection{RQ2: Performance in Post-learning Phase}\label{sec.results.rq2}
\looseness-1To address RQ2, we assess the performance of the groups in the post-learning phase on \curradv{}. The distribution of scores on \curradv{} for each learning group is shown in Figure~\ref{fig5:overall.hocmaze}. In Figure~\ref{fig5:overall.aggr}, we present the normalized aggregated performance (scaled between $0$ and $1$) of both groups on the post-learning phase items (P01--P15) from \curradv{}. A $\chi^2$ significance analysis~\citep{cochran1952chi2} showed that \currtwo{} significantly outperformed \currone{} in aggregate performance ($\chi^2 = 75.1, p < 0.01$). $\chi^2$ analysis w.r.t. each item category also showed that \currtwo{} significantly outperformed \currone{} (all $p$-values $< 0.01$). Our analysis highlights that the group with richer quiz types better transferred their problem-solving skills in the post-learning phase compared to the group with basic quizzes.

\looseness-1Next, as discussed in Section~\ref{sec.study.analysis.setup}, we also compare the performance of learning groups to the baseline performance on \curradv{} to further evaluate the relative utility of quizzes. Interestingly, the difference $\Delta$ in the aggregate performance of the groups ($0.799$ and $0.702$) w.r.t. baseline performance ($0.763$) is as follows: $\Delta = \textcolor{black}{+0.036}$ for \currtwo{} and $\Delta = \textcolor{black}{-0.061}$ for \currone{}. In fact, on the challenging items P10--P12 unique to \curradv{}, the difference in performance is larger: $\Delta = \textcolor{black}{+0.081}$ for \currtwo{} and $\Delta = \textcolor{black}{-0.068}$ for \currone{}. The finding that \currone{} performed worse than the baseline performance, suggests an overreliance on basic quizzes during learning, and requires further investigation into why basic quizzes might negatively impact long-term performance~\citep{deslauriers2019measuring,karpicke2011retrieval}.

\begin{tcolorbox}
\textbf{Answer to RQ2}: The experimental group significantly outperforms the control group in the post-learning phase on \curradv{}.
\end{tcolorbox}

\subsection{RQ3: Impact of Prior Experience}\label{sec.results.rq3}
To address RQ3, we investigate which learning group performs better when considering the students' grades (year of study) and years of prior programming experience. Aggregated normalized performance on \curradv{}, segmented by grade and years of programming experience, are illustrated in Figure~\ref{fig6:overall}. A $\chi^2$ significance analysis~\citep{cochran1952chi2} showed that \currtwo{} significantly outperformed \currone{} across both grades. Interestingly, the performance gap between the two groups is higher when considering the lower grade. Furthermore, when accounting for students' prior programming experience, we find that \currtwo{} significantly outperformed \currone{} across all categories. Overall, our analysis highlights the utility of using richer quizzes compared to basic quizzes across learners with varying experiences.

\begin{tcolorbox}
\textbf{Answer to RQ3}: The experimental group outperforms the control group on \curradv{} among all categories of students, with higher benefits for students in the lower grade.
\end{tcolorbox}

\subsection{Limitations}\label{sec.results.limitations}
\looseness-1Next, we discuss a few limitations of our study. First, we did not use an explicit pre-test to measure pre-knowledge across the learning groups. We could not reuse intricate write-code tasks as a pre-test due to potential negative effects on motivation; moreover, we did not use other quizzes in the pre-test as that might interfere with evaluating our quiz-based curricula. In the future, one could reduce discrepancies in pre-knowledge across groups by scaling up the study and possibly doing randomization at the student level (instead of class level). Second, the sequence of items in both groups during learning was decided based on discussions with teachers and a pilot study. In the future, one could conduct rigorous studies to optimize the sequence. Finally, the $41$ items for each group during the learning phase were spread across two $45$-minute lessons over two weeks, complicating comparisons with the standard \currhoc{} curriculum of $20$ items that are typically covered in one $45$-minute lesson. In the future, it would be useful to design a quiz-based curriculum that can be covered in one lesson (e.g., by selecting fewer write-code tasks and quizzes), allowing for a more direct comparison with a baseline curriculum with only write-code tasks.

\section{Concluding Discussions}\label{sec.conclusion}
\looseness-1We compared two learning groups incorporating different types of quizzes interleaved with write-code visual programming tasks for K-$8$ learners. This comparison was conducted within the widely used \emph{Hour of Code: Maze Challenge} by code.org for elementary programming. The experimental group had richer quizzes, including code debugging and task design, while the control group had basic solution-finding quizzes. Our findings show that the experimental group significantly outperformed the control group in the post-learning phase, despite performing similarly in the learning phase.

There are many interesting directions for future work. First, additional think-aloud studies could be conducted to gain insights into engagement patterns with different quiz types. This could help explain the over-reliance on basic quizzes during the learning phase and its resulting negative transfer in the post-learning phase. Second, it would be interesting to extend the investigation of quiz-based exercises to more advanced programming contexts like Python. Finally, it would be useful to develop an adaptive scaffolding framework centered around these richer quizzes, where they are presented to learners as and when they seek assistance.

\begin{acks}
We would like to thank the teachers and students in Estonia for their participation in the study. Funded/Cofunded by the European Union (ERC, TOPS, 101039090). Views and opinions expressed are however those of the author(s) only and do not necessarily reflect those of the European Union or the European Research Council. Neither the European Union nor the granting authority can be held responsible for them.
\end{acks}

\bibliographystyle{ACM-Reference-Format}
\balance
\bibliography{main}

\end{document}